\documentclass[11pt]{article}

\usepackage{amsmath,amsthm,amssymb}
\usepackage{graphicx,color}
\usepackage{setspace}
\usepackage{hyperref}

\newcommand{\Z}{\mathbb{Z}}

\newcommand{\QC}{quasicrystal}
\newcommand{\QCn}{quasicrystalline}

\makeatletter
\newcommand{\ps@paper}{%
  \renewcommand{\@oddhead}{%
    {\itshape 4-Coordinated Quasicrystal}\hfill \thepage}%
  \renewcommand{\@evenhead}{%
    {\itshape 4-Coordinated Quasicrystal}\hfill \thepage}%
  \renewcommand{\@oddfoot}{}%
  \renewcommand{\@evenfoot}{}}
\makeatother

\theoremstyle{plain}

\theoremstyle{definition}

\theoremstyle{remark}

\begin{document}


\title{An Icosahedral Quasicrystal as a Packing of Regular Tetrahedra}

\author{F.\ Fang\thanks{Corresponding author. Email: {\tt fang@quantumgravityresearch.org}},\hspace{.5ex} J.\ Kovacs, G.\ Sadler, and K.\ Irwin\\[8pt]{\em Quantum Gravity Research,} Los Angeles, CA, U.S.A.}

\date{}

\maketitle

\pagestyle{plain} 
\setcounter{page}{1}

\begin{abstract}
We present the construction of a dense, \QCn\ packing of regular tetrahedra with icosahedral symmetry. This \QCn\ packing was achieved through two independent approaches. The first approach originates in the Elser-Sloane 4D \QC\ \cite{Elser1987}. A 3D slice of the \QC\ contains a few types of prototiles. An initial structure is obtained by decorating these prototiles with tetrahedra. This initial structure is then modified using the Elser-Sloane \QC\ itself as a guide. The second approach proceeds by decorating the prolate and oblate rhombohedra in a 3-dimensional Ammann tiling \cite{Peters1991}. The resulting \QC\ has a packing density of 59.783\%. We also show a variant of the \QC\ that has just 10 `plane classes' (compared with the 190 of the original), defined as the total number of distinct orientations of the planes in which the faces of the tetrahedra are contained. This small number of plane classes was achieved by a certain `golden rotation' of the tetrahedra \cite{Fang2013}.
\end{abstract}

\section*{Introduction}
The emergence of ordered structure as a result of self-assembly of building blocks is far from being well understood, and has received a great deal of attention \cite{Phillips2012a,Phillips2012b,Damasceno2012}. Understanding this underlying scheme (an example of `hidden order') may open up the doors for explaining phenomena that has thus far remained elusive (e.g., anomalies in water \cite{Brovchenko2008}). Studies have shown that order in structure is directly related to its fundamental building blocks \cite{Damasceno2012}. Recently, 2D \QCn\ order has been achieved in a tetrahedral packing, when thermodynamic conditions are applied to an ensemble of tetrahedra \cite{Akbari2009}. This discovery has stimulated further investigations into \QCn\ packings of tetrahedra. This paper presents an icosahedrally symmetric \QC, as a packing of regular tetrahedra. We obtain this by using two approaches, with very different rationales, but ultimately obtain the same \QCn\ packing. This structure might also help in the search for a perfect 4-coordinated \QC\ \cite{Dmitrienko2001}: even though the tetrahedra in our \QC\ are packed in such a way that faces of neighboring tetrahedra can be paired up (yielding a 4-connected network between tetrahedra centers), the structure cannot be considered physically as 4-coordinated, due to the excessively high ratio between the longest and shortest bonds stemming from each tetrahedron center (Figure~\ref{network}~{\bf b}).

\section*{First approach}
\begin{figure}[h!]
  \centering
  \includegraphics[width=\textwidth,clip]{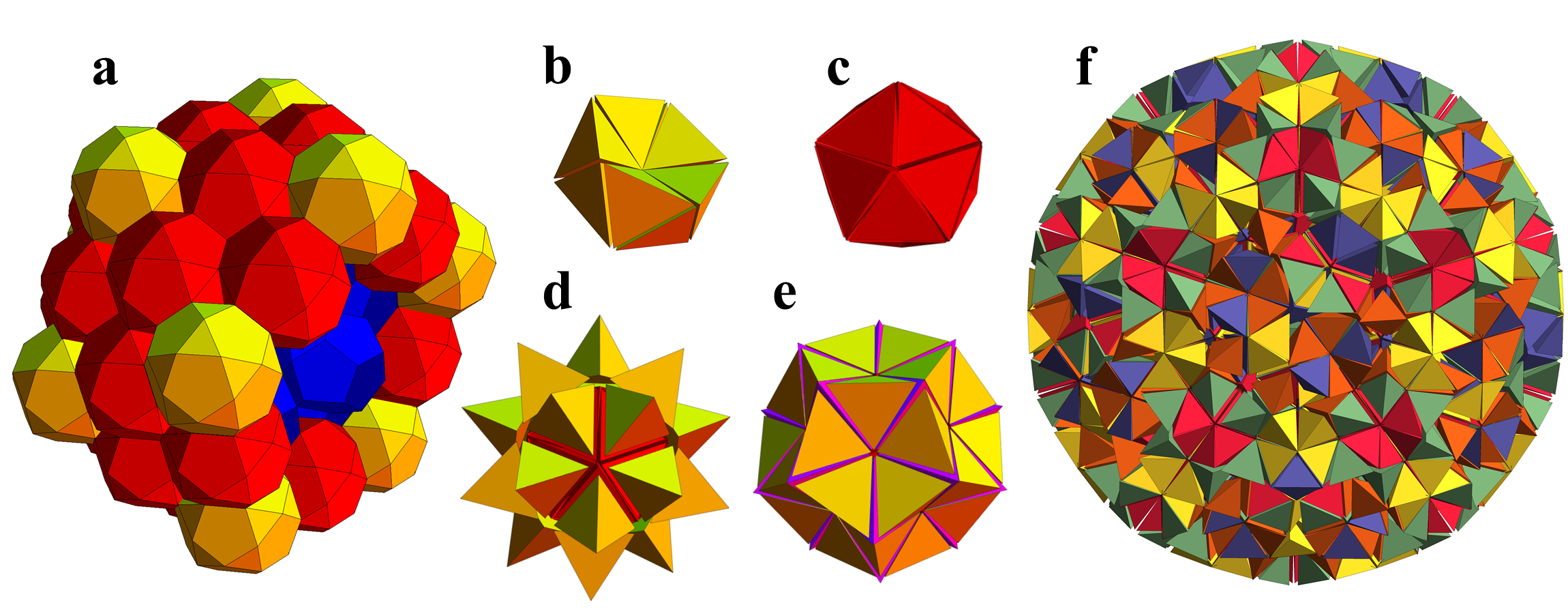}
  \caption{({\bf a}) 3D slice of the Elser-Sloane \QC. ({\bf b}) 10-tetrahedron `ring'. ({\bf c}) 20-tetrahedron `ball' (icosahedral cluster), obtained by capping the ring with two 5-tetrahedron groups. ({\bf d}) 40-tetrahedron cluster, obtained by placing a tetrahedron on top of each face of the icosahedron in ({\bf c}). ({\bf e}) 70-tetrahedron dodecahedral cluster, obtained by adding 30 more tetrahedra in the crevices in ({\bf d}). ({\bf f}) Patch of the resulting \QC, which contains the clusters shown in ({\bf c}), ({\bf d}) and ({\bf e}) around its center.}
  \label{clusters}
\end{figure}

This is a guided decoration of the 3D slice of the Elser-Sloane \QC\ \cite{Elser1987} (Figure~\ref{clusters}~{\bf a}), which contains four types of prototiles: icosidodecahedron (IDHD), dodecahedron (DHD), and icosahedron (IHD)---each as a section (boundary of a cap) of a 600-cell---and a golden tetrahedron. The way each of these polyhedra (Figure~\ref{clusters}~{\bf c}-IHD, {\bf e}-DHD, IDHD (not shown)) is decorated is guided by the arrangement of the tetrahedra in the 600-cells of the Elser-Sloan \QC. For example, the IDHD is a slice through the equator of the 600-cell and is the boundary of a cap of 300 tetrahedra. Projecting these 300 tetrahedra into the IDHD hyperplane results in distorted tetrahedra, and gaps appear when tetrahedra are restored (while avoiding collisions) back to a regular shape. To maintain a higher packing density and a better tetrahedral coordination \cite{Dmitrienko2001}, extra tetrahedra can be introduced to fill in such gaps or, alternatively, some tetrahedra may be removed from each shell before regularization to avoid conflicts resulting from collisions. Both methods result in exactly the same \QCn\ structure, as shown in Figure~\ref{clusters}~{\bf f}. This method can be thought as a decoration of the 3D slice of the Elser-Sloane \QC, guided by the \QC\ itself.

\section*{Second approach}
This is an direct decoration of the 3D Ammann tiling \cite{Peters1991}. A 3D Ammann tiling can be generated by cut-and-project from the $\Z^6$ lattice, and contains two prototiles: a prolate rhombohedron and an oblate rhombohedron.

\begin{figure}[h!]
  \centering
  \includegraphics[width=.54\textwidth,clip]{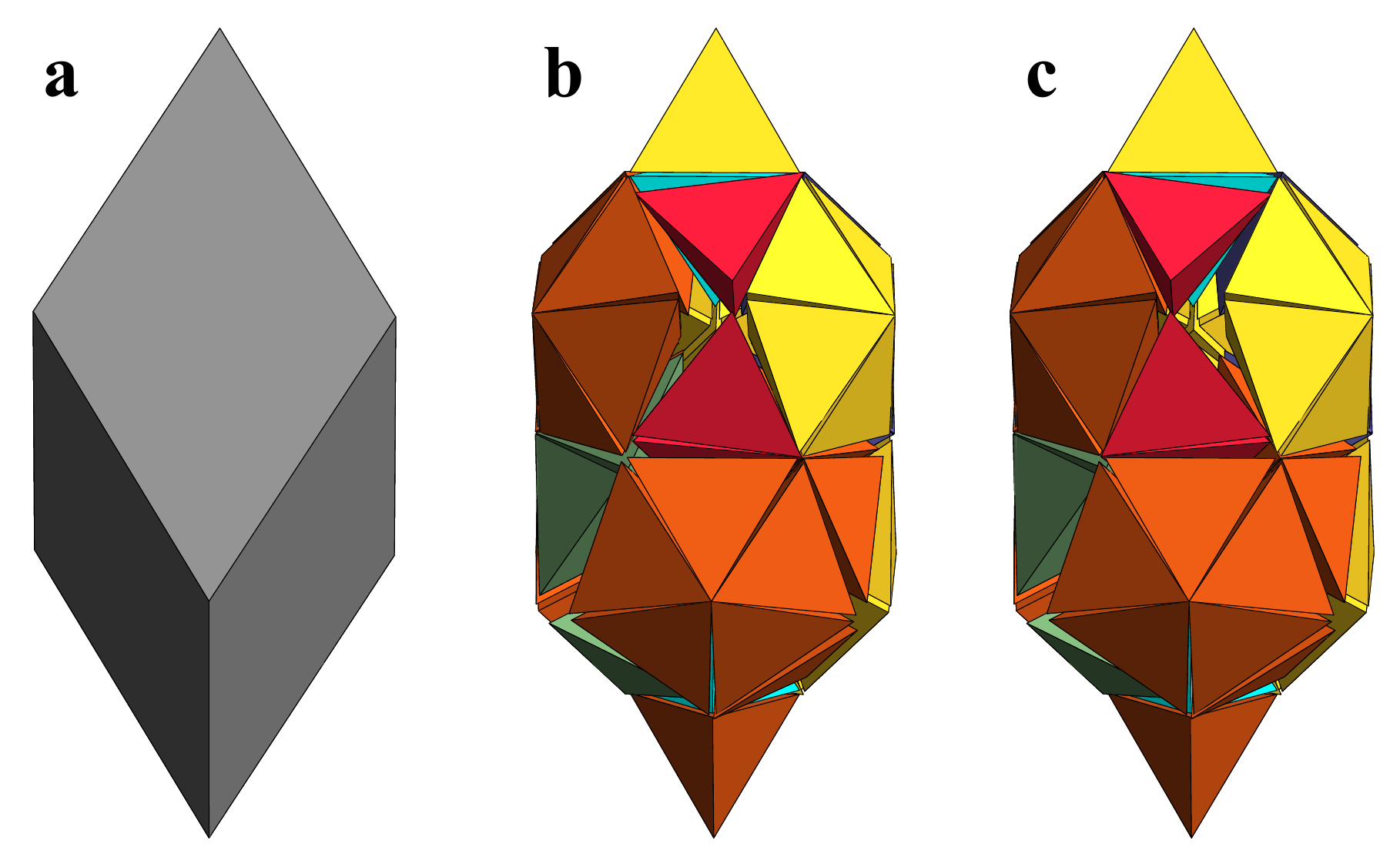}
  \caption{Decoration of the prolate rhombohedron ({\bf a}). ({\bf b}), ({\bf c}) The two conformations that 3 of tetrahedra associated to each face can have. There are 52 tetrahedra decorating the whole prolate rhombohedron: 16 of them lie completely inside it, while the remaining 36 are shared 50\% with corresponding tetrahedra in a face of an adjacent rhombohedron in the packing. The 3 tetrahedra (in each face) that can `flip' consist of 2 shared ones and one internal one.}
  \label{prolate}
\end{figure}

\begin{figure}[h!]
  \centering
  \includegraphics[width=.75\textwidth,clip]{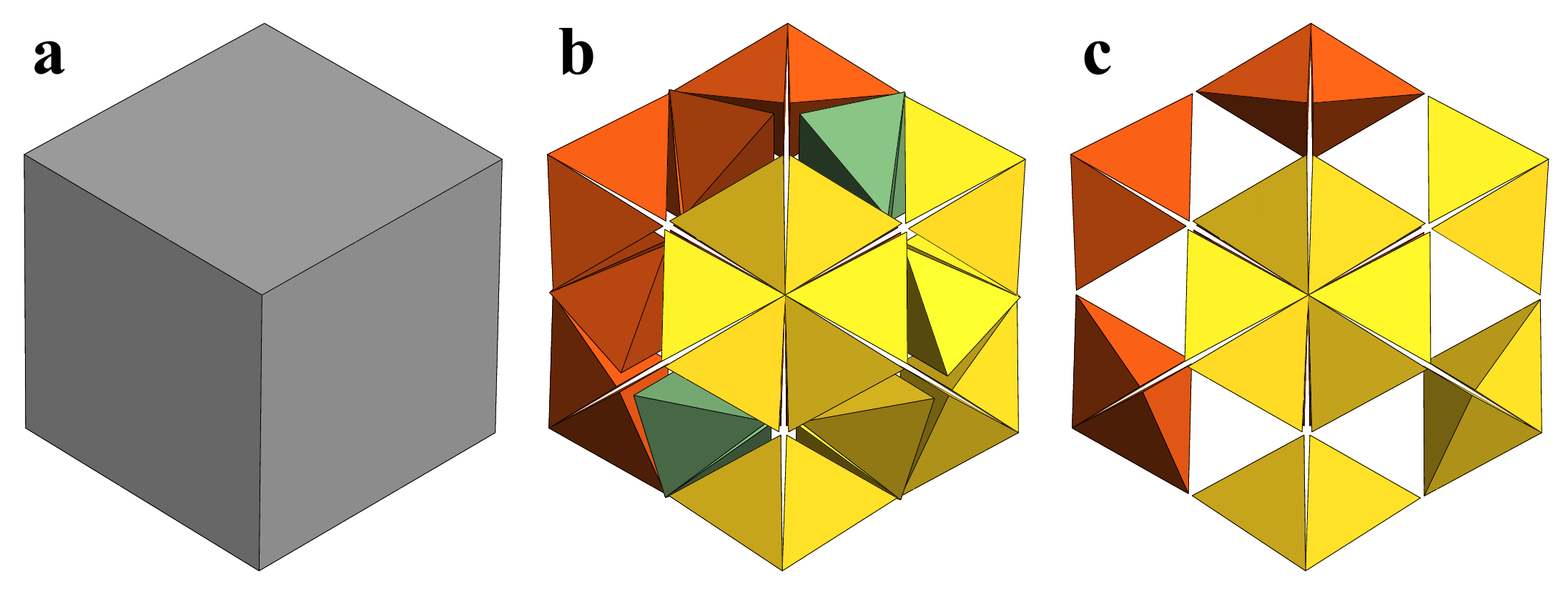}
  \caption{Decoration of the oblate rhombohedron ({\bf a}). ({\bf b}) The decoration consists of 36 tetrahedra, each shared 50\% with corresponding tetrahedra in an adjacent rhombohedron. ({\bf c}) Here only the 24 tetrahedra coming from the 20-tetrahedron `balls' (each centered at a vertex of the rhombohedron) are displayed, in order to get a better feel of the arrangement. The other 12 tetrahedra shown in ({\bf b}) come from the 10-tetrahedron `rings' (each centered at the midpoint of a rhombohedron edge).}
  \label{oblate}
\end{figure}

The decoration consists in placing a 20-tetrahedron `ball' (Figure~\ref{clusters}~{\bf c}) at each vertex, and a 10-tetrahedron `ring' (Figure~\ref{clusters}~{\bf b}) around each edge of each of the rhombohedra, and then removing all those tetrahedra that do not intersect the rhombohedron. This process creates some clashes inside the oblate rhombohedron, and after excluding the appropriate tetrahedra, the packing remains face-to-face and therefore the resulting network of tetrahedron centers is 4-connected. (However, the bond-length distribution has a rather long tail, (Figure~\ref{network}~{\bf b}).) This process yields some pairs of tetrahedra with large overlaps shared between the balls and the rings, providing degrees of freedom to choose either tetrahedron of each pair, which in turn translates into the ability to flip 3 tetrahedra in each face of the prolate rhombohedra, suggesting a novel phason mechanism for this type of \QC\ (Figure~\ref{prolate}~{\bf b,c}). The oblate rhombohedron, due to its flatness, does not have these degrees of freedom, there being only one way to choose the tetrahedra so that the resulting decoration will have a 3-fold axis of symmetry (Figure~\ref{oblate}).

\section*{Twisting and plane-class reduction}
\begin{figure}[h!]
  \centering
  \includegraphics[width=.75\textwidth,clip]{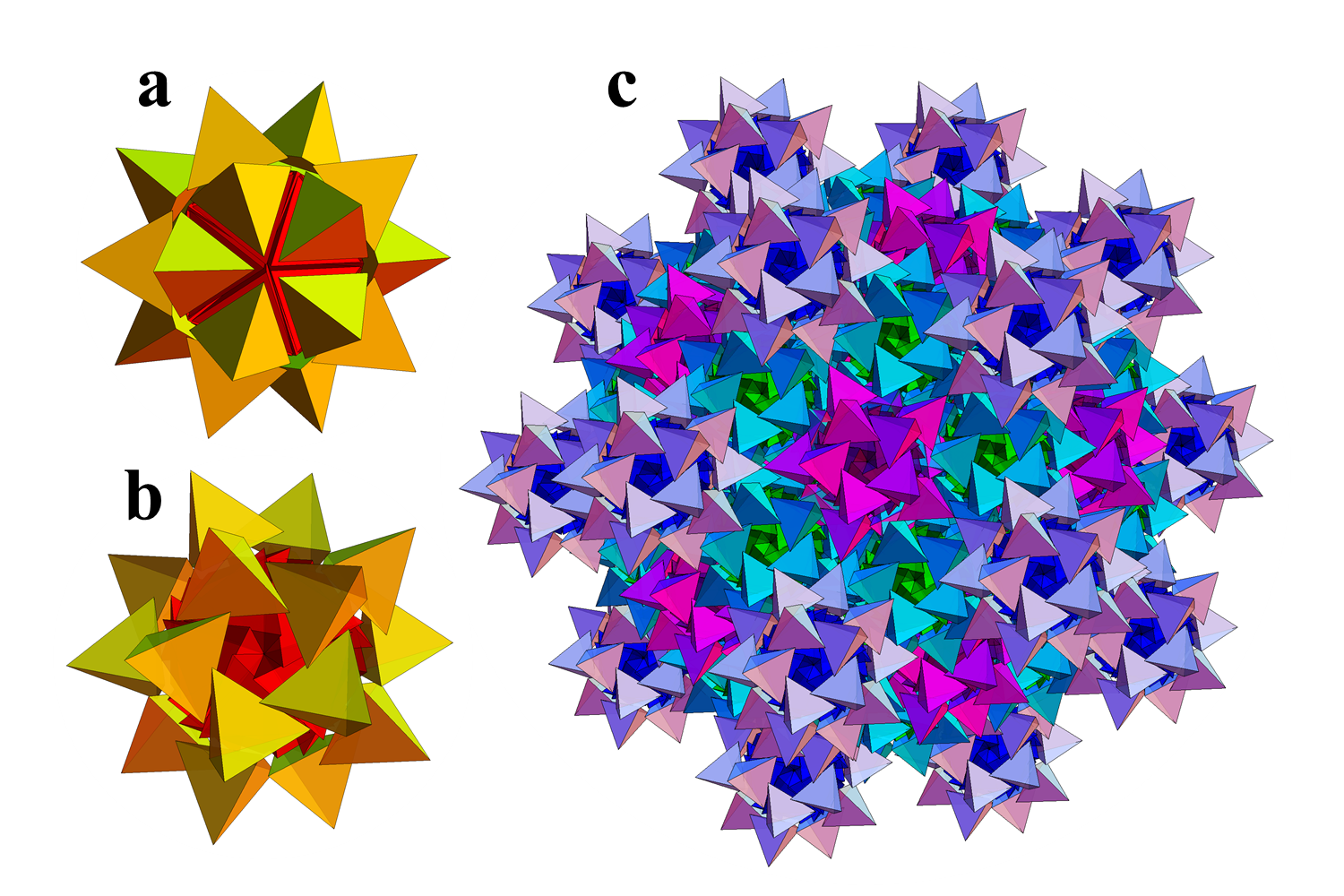}
  \caption{({\bf a}) Non-twisted and ({\bf b}) twisted 40-tetrahedron clusters. ({\bf c}) Patch of the twisted \QC.}
  \label{twisted}
\end{figure}

The \QC\ can also be obtained by placing a 40-tetrahedron cluster (Figure~\ref{twisted}~{\bf a}) at each vertex of the 3D Ammann tiling. Applying a `golden rotation' \cite{Fang2013} of $\arccos(\tau^2/2\sqrt{2}) \approx 22.2388^{\circ}$ (where $\tau=\frac{1}{2}(1+\sqrt{5})$ is the golden ratio) to each of the tetrahedra in the cluster, around an axis running through the tetrahedral center and the center of the cluster, yields a twisted \QC\ (Figure~\ref{twisted}~{\bf b}-twisted 40-tetrahedron cluster, {\bf c}-the twisted \QC). This golden twist reduces the total number of plane classes from 190 to 10. The resulting relative face rotation at each `face junction' between adjacent tetrahedra is $\arccos(\frac{1}{4}(3\tau-1)) = \frac{1}{3}\arccos(11/16) \approx 15.5225^{\circ}$.

\section*{Analysis}
\begin{figure}[h!]
  \centering
  \includegraphics[width=.9\textwidth,clip]{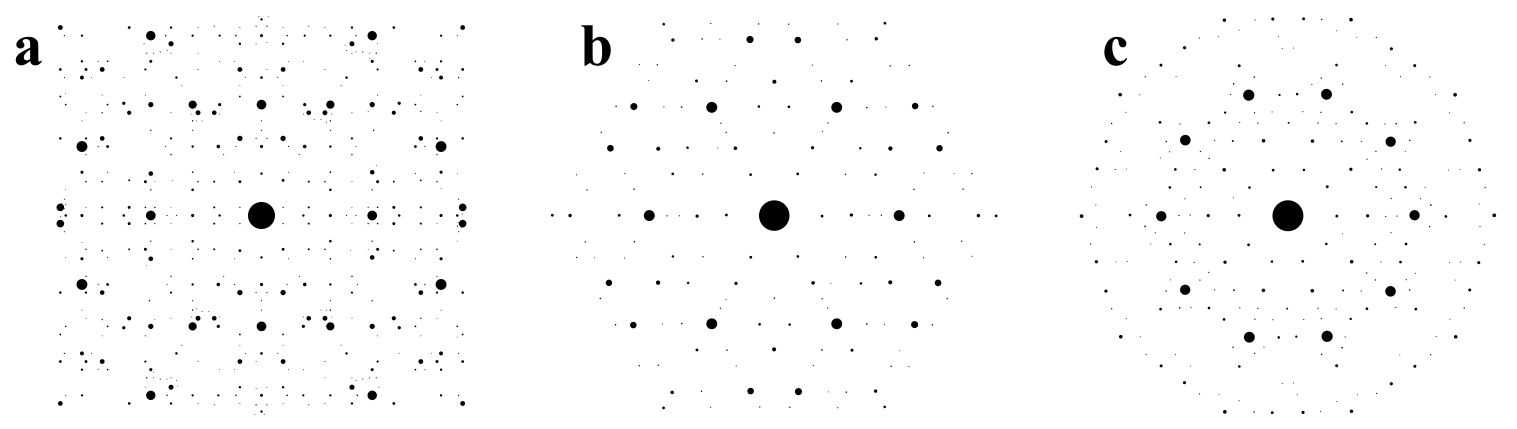}
  \caption{Diffraction patterns on the 2-fold ({\bf a}), 3-fold ({\bf b}), and 5-fold ({\bf c}) planes of the \QC\ defined by the tetrahedral centers.}
  \label{diffract}
\end{figure}

Diffraction patterns of the non-twisted \QC\ (Figure~\ref{clusters}~{\bf f}) reveal 2-, 3-, and 5-fold symmetry planes (Figure~\ref{diffract}), confirming the icosahedral symmetry of this \QC. Its packing density is $\frac{65}{16464}(208800\sqrt{2}+64215\sqrt{5}-45499\sqrt{10}-294845) \approx 0.59783$. The derivation of this expression is too lengthy to be included here, and was done using the {\sl Mathematica} software. (A {\sl Mathematica} notebook containing code for this calculation is available upon request.) Basically, the calculation is done for each rhombohedron, by solving equations that minimize the rhombohedron's edge length relative to the tetrahedron's, in such a way that the various tetrahedra just touch each other. The degrees of freedom allowed in this process are shifts in the directions of the rhombohedron's edges and radially from them. Finally, the density values for both rhombohedra are combined using the fact that the relative frequencies of occurrence of the prolate and oblate rhombohedra in the Ammann tiling is the golden ratio.

\begin{figure}[h!]
  \centering
  \includegraphics[width=.9\textwidth,clip]{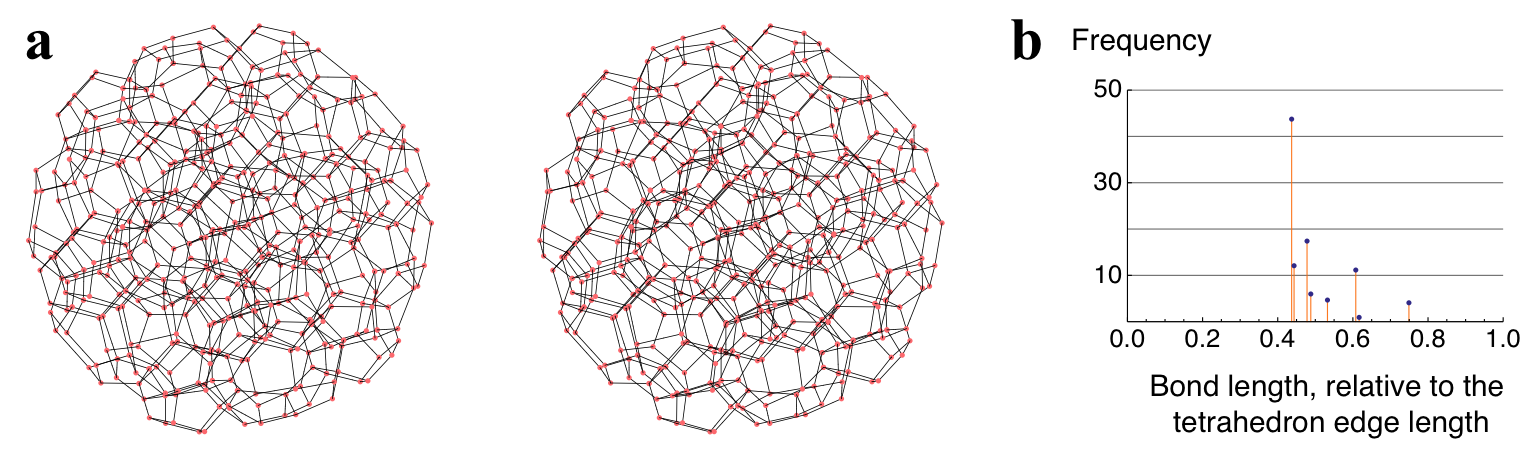}
  \caption{({\bf a}) Stereo pair of the network of tetrahedral centers. ({\bf b}) Distribution of bond lengths in the network.}
  \label{network}
\end{figure}

The network of tetrahedral centers (Figure~\ref{network}~{\bf a}) is 4-connected, although it cannot be considered what the chemistry community would call ``4-coordinated,'' due to the rather wide range of bond lengths (Figure~\ref{network}~{\bf b}), with a ratio of 1.714 between the longest and shortest bonds.

\section*{Summary and outlook}
Using two seemingly unrelated approaches, we have surprisingly obtained the same \QCn\ packing of regular tetrahedra with global icosahedral symmetry. After the fact, this convergence turned out not to be by chance. The reason is that a 3-dimensional slice (in the appropriate orientation) of the 4-dimensional Elser-Sloane \QC\ can be obtained directly by cut-and-project from the $D_6$ lattice \cite{Moody1993}, $D_6$ being a sublattice of $\Z^6$ (and $\Z^6$ being a sublattice of $\frac{1}{2}D_6$, the 6-dimensional face-centered cubic lattice).

To our knowledge, an icosahedrally symmetric packing of tetrahedra with this relatively high density of almost 0.6 has not been shown before. Moreover, this packing provides a 4-connected network with bond lengths ranging from 0.437 and 0.749 (in units of the tetrahedron's edge length), a 1:1.714 ratio. These features are non-trivial among icosahedral arrangements of tetrahedra. For applications, an important step would be to shrink the range of bond lengths to what can be considered as a realistic 4-coordinated network.

We also considered the number of `plane classes' and ways to reduce it to the minimum possible. The above packing has 190 plane classes. By applying what we call the `golden twist' to each tetrahedron, the 190 plane classes of the original \QC\ coalesce to only 10. (This is easily seen to be the minimum possible for an icosahedral arrangement of tetrahedra.)

This \QC\ also suggests interesting alternatives to the classical phason flips, as shown in Figure~\ref{prolate}. We are investigating the dynamics of this and other types of phasons and their potential physical applications.

\bibliographystyle{abbrv}
\bibliography{m-science}

\end{document}